\documentstyle{article}
\def\ra{{\rm rank}}
\def\pa{\partial}
\def\de{\delta_\epsilon}
\def\ea{\epsilon^a}
\title{Notes on Lagrangean and Gamiltonian Symmetries}
\author{A. A. Deriglazov\thanks{E-mail: theordpt@fftgu.tomsk.su}}
\date{Department of Mathematical Physics,\\
Tomsk Polytechnical University,\\
Tomsk 634004, Russia}
\begin{document}
\maketitle
\begin{abstract}
The Hamiltonization of local symmetries of the form $\delta
q^A=\ea{R_a}^A(q,\dot q)$ or $\delta q^A=\dot\ea{R_a}^A
(q,\dot q)$ for arbitrary Lagrangean model $L(q^A,\dot q^A)$ is
considered. We show as the initial symmetries are transformed in the
transition from $L$ to first order action, and then to the Hamiltonian
action $S_H=\int{\rm d}\tau(p_A\dot q^A-H_0-v^\alpha\Phi_\alpha)$,
where $\Phi_\alpha$ are the all (first and second class) primary
constraints. An exact formulae for local symmetries of $S_H$ in terms
of the initial generators ${R_a}^A$ and all primary constraints
$\Phi_\alpha$ are obtained.
\end{abstract}
\newpage
\section{Introduction}

In the majority of physically interesting models, the Lagrangians are
symmetric with respect to some set of local transformations of the form
$$
\de q^A=R^A(\ea,q^A,\dot q^A).
\eqno{(1)}$$
On application of the Dirac--Bergmann algorithm \cite{1, 2} for the
Hamiltonization of the theory under investigation, we obtain an
equivalent description for the original classical dynamics in terms of
canonical action
$$
S_c=\int{\rm d}\tau\left(-{1\over 2}C^{-1}_{AB}\dot\Gamma^A\Gamma^B-
H_0-v^I\Phi_I\right),
\eqno{(2)}$$
where $\Gamma^A\equiv(q^A,p_B)$ and $\Phi_I(q,p)$ are all first-class
constraints of the theory. Constraints of the second class, if any, are
taken into account by going to the Dirac bracket, such that
$\{\Gamma^A,\Gamma^B\}=C^{AB}$. It is well known \cite{3, 4} that local
symmetries for Eq. (2) are the following transformations generated (in
the sector $(q^A,p_B)$) by the first-class constraints $\Phi_I$:
$$
\delta q^A=e^I\{q^A,\Phi_I\},
$$
$$
\delta p_A=e^I\{p_A,\Phi_I\},
\eqno{(3)}$$
$$
\delta v^I=\dot\epsilon^I+v^I\epsilon^K{C_{JK}}^I-\epsilon^J{V_J}^I,
$$
where the designations $\{\Phi_J,\Phi_K\}={C_{JK}}^I\Phi_I$,
$\{H_0,\Phi_I\}={V_I}^J\Phi_J$.

For any specific model, numerous observations are available on how the
Lagrangian symmetries, Eq. (1), and the Hamiltonian symmetries, Eq.
(3), are bound \cite{5, 6} but the question about the relation between
them in a general theory still remains largely open. In particular,
some researchers [3, 7--10] investigated the problem of
reconstruction of Lagrangian symmetries by known Hamiltonian
symmetries. However, provided that the Dirac--Bergmann algorithm is
applied, the most natural statement of the problem seems to be as
follows: how the Lagrangian transformations, Eq. (1), change when
passing successively from the Lagrangian to the canonical action? In
other words, concurrent with the procedure of Hamiltonization of the
theory, we state the problem of Hamiltonization for the Lagrangian
transformations with the aim to obtain an expression for the
Hamiltonian action symmetries through the generators
$R^A(\epsilon,q,\dot q)$.

The solution of this problem may appear to be useful for a number of
issues, in particular, in studying the corresponding algebras
\cite{11}; in investigating the relation between the Lagrangian and the
Hamiltonian BFV quantization \cite{12}; in consistent formulating the
theory of the superparticle (superstring) on a curved background
\cite{13}, and discussing Dirac's conjecture \cite{8, 14}.

The present work is organized in the following way. In Sec. 2, we
introduce designations and give the facts related to the
Dirac--Bergmann algorithm that are necessary for the subsequent
discussion (see Ref. 2 for details).In Sec. 3, the symmetries for the
action in the first-order formalism $S_v$ and for the Hamiltonian
action $S_H$ are constructed by the local symmetries of the original
action. Subsequently, a partial Hamiltonization of the transformations
is performed for both cases, i.e., they are rewritten in terms of the
Poisson bracket to the trivial (on-shell vanishes) symmetries of the
first-order formalism. Section 4 deals with a special case of original
Lagrangian symmetries where a ``complete'' Hamiltonization appears to
be possible, namely, the symmetries $S_H$ are expressed through all
primary constraints of the theory (and through the generators of the
original Lagrangian symmetries). It should be noted that this special
class of symmetries is rather broad; in particular, the local fermion
symmetries of the superparticle and the superstring theories in a
covariant formulation satisfy the restrictions placed. The results are
formulated and discussed in Conclusion.

\section{The Dirac--Bergmann algorithm}

Consider a mechanical system described by at most polynomial in
velocities Lagrangian $L(q^A,\dot q^A)$, $A=1,\dots,N$, which will be assumed
singular:
$$
\ra\,M_{AB}\equiv\ra\frac{\pa^2L}{\pa\dot q^A\pa\dot q^B}=K<N.
\eqno{(4)}$$
According to this equation, it is convenient to reliable the index
$A\equiv(i,\alpha)$, $i=1,\dots,K$, $\alpha=K+1,\dots,N$, such that
$M_{ij}$ is non-singular and its inverse $\tilde M^{ij}$ exists (note
that for all known models this may be done without losing of manifestly
Poincar\'e covariance). We also suppose that all variable $q^A$ are
even, the extension of all results to grassmanian case is formally
straightforward. Further, let us consider an infinitesimal local
transformations of the form
$$
\de q^A=\ea{R_a}^A(q^B,\dot q^B), \qquad
a=1,\dots,k',
\eqno{(5)}$$
and suppose the $L$ is invariant up to an exact differential
$$
\de L=[\ea D_a(q,\dot q)]^..
\eqno{(6)}$$
If Eq. (5) essentially depends on the parameters $\ea(\tau)$
($\ra\,{R_a}^A=\max$), then $K'\leq N-K$ as it will be seen from Eq. (27)
below.

To go from the Lagrangian to the Hamiltonian formalism we first to
pass into equivalent description of the initial dynamics in terms of
first order action, defined on extended space $(q^A,p_A,v^A)$
$$
S_v=\int{\rm d}\tau[L(q^A,v^B)+p_A(\dot q^A-v^A)].
\eqno{(7)}$$
The equations of motion which follows from Eq. (7) may be identically
rewritten in the Hamiltonian form by introducing of the Poisson bracket
\{ , \} (defined only for phase space sector $(q^A,p_A)$ of extended
space), and of the Hamiltonian
$$
\bar H(q^A,p_A,v^A)\equiv p_Av^A-L(q,v).
\eqno{(8)}$$
Then the dynamics is ruled by the following equations:
$$
\dot q^A=\{q^A,\bar H\},
\eqno{(9)}$$
$$
\dot p^A=\{p_A,\bar H\},
\eqno{(10)}$$
$$
\Phi_\alpha(q,p,v)\equiv p_\alpha-{\pa L\over\pa v^\alpha}=0,\quad
\mbox{or}\quad {\pa\bar H\over\pa v^\alpha}=0,
\eqno{(11)}$$
$$
\Phi_i(q,p,v)\equiv p_i-{\pa L\over\pa v^i}=0,\quad
\mbox{or}\quad {\pa\bar H\over\pa v^i}=0.
\eqno{(12)}$$

As a second step of Dirac--Bergmann algorithm, we solve Eq. (12)
$$
v^i=v^i(q^A,p_j,v^\alpha),
\eqno{(13)}$$
and substitute these back into (9)--(12). Then we have the identities
$$
p_i-{\pa L\over\pa v^i}\biggm|_v\equiv 0, \quad \mbox{or}\quad
{\pa\bar H\over\pa v^i}\biggm|_v\equiv 0,
\eqno{(14)}$$
and the equations of motion in reduced space $(q^A,p_A,v^\alpha)$
$$
\dot q_A=\{q^A,H\},
\eqno{(15)}$$
$$
\dot p_A=\{p_A,H\},
\eqno{(16)}$$
$$
\Phi_\alpha(q^A,p_A)\equiv p_\alpha-{\pa L\over\pa
v^\alpha}\biggm|_v=0,
\eqno{(17)}$$
where
$$
H(q^A,p_A,v^\alpha)\equiv\bar H\bigm|_v.
\eqno{(18)}$$

Note that $\{A(q,p),\bar H\}\bigm|_v\equiv\{A(q,p),\bar
H\bigm|_v\}$ as a consequence of Eq. (14). Also, the left hand side
of Eq. (17) do not depend on $v^\alpha$, in accordance with the
condition (4).

It is well known that the Hamiltonian $H$ may be identically rewritten
in the form
$$
H=H_0(q^A,p_i)+v^\alpha\Phi_\alpha,
\eqno{(19)}$$
where
$$
H_0\equiv\left(p_iv^i-L(q,v)+v^\alpha{\pa L\over\pa v^\alpha}\right)
\biggm|_v,
\eqno{(20)}$$
the last may depend only on $q^A$ and $p_i$ variables.

As the result, the Hamiltonian dynamics may be described in terms of
Hamiltonian action
$$
S_H\equiv S_v\bigm|_v=\int{\rm d}\tau(p_A\dot q^A-H_0-v^\alpha
\Phi_\alpha),
\eqno{(21)}$$
where $\Phi_\alpha$ are the all (first and second class) primary
constraints. Equations (15)--(17) follows from variation of Eq. (21)
with respect to $q,p,v$ variables.

In conclusion of this section let us write some identities will be used
below. By differentiating of Eq. (14) one get
$$
{\pa v^i\over\pa p_A}=\tilde M^{ij}\bigm|_v{\delta_j}^A,
\eqno{(22)}$$
$$
{\pa v^i\over\pa v^\alpha}=-\tilde M^{ij}M_{j\alpha}\bigm|_v,
\eqno{(23)}$$
$$
{\pa v^i\over\pa q^A}=-\tilde M^{ij}{\pa^2L\over\pa v^j\pa q^A}
\biggm|_v,
\eqno{(24)}$$
Where, from now $M_{AB}\equiv\pa^2L(q,v)/\pa v^A\pa v^B$. Then, from
the identity \linebreak $\pa\bigl(\Phi_\alpha\bigm|_v\bigr)/\pa v^\beta\equiv
0$ and
from Eq. (23) one finds
$$
(M_{\alpha\beta}-M_{\alpha i}\tilde M^{ij}M_{j\alpha}\bigm|_v\equiv
0.
\eqno{(25)}$$

\section{Hamiltonization of local symmetries}

To rewrite the local symmetries of $L$ in terms of $S_v$ and $S_H$, let
us first to arrive at a consequences, followed from the condition
(6). By standard algebraic manipulations, one finds an expression for
an exact differential
$$
{\pa L\over \pa\dot q^A}{R_a}^A=D_a,
\eqno{(26)}$$
and the following Lagrangian identities
$$
\begin{array}{c}\displaystyle\frac{\pa^2L}{\pa\dot q^A\pa\dot q^B}
{R_a}^A(q,\dot q)=0,\\[12pt]
\displaystyle\frac{\pa L}{\pa q^A\pa\dot q^B}{R_a}^A-
\displaystyle\frac{\pa L}{\pa q^B\pa\dot q^A}\dot q^B{R_a}^A=0.
\end{array}\eqno{(27)}$$
As it fulfilled for arbitrary $q^A(\tau)$ these equations remain valid
after the substitution $\dot q^A\to v^A$ and identically fulfilled for
arbitrary functions $q^A(\tau)$ and $v^A(\tau)$.

Now, by using Eqs. (26) and (27), one may easily check that following
local transformations
$$
\de q^A=\ea{R_a}^A(q,v),
$$
$$
\de p_A={\pa^2L\over\pa q^A\pa v^B}\de q^B,
\eqno{(28)}$$
$$
\de v^A=(\de q^A)^.
$$
leaves the action $S_v$ (7) invariant up to boundary terms. Further,
these formulae may be identically expressed in terms of Poisson
brackets as follows:
$$
\de q^A=\ea\{q^A,\Phi_B{R_a}^B\},
\eqno{(29)}$$
$$
\de p_A=\ea\{p_A,\Phi_B{R_a}^B\}+\Phi_B\ea
{\pa{R_a}^B\over\pa q^A},
\eqno{(30)}$$
$$
\de v^A=\dot\epsilon{R_a}^A+\ea\left({\pa{R_a}^A
\over\pa q^B}v^B+{\pa{R_a}^A\over\pa v^B}\dot v^B\right)+\ea
{\pa{R_a}^A\over\pa q^B}(\dot q^B-v^B),
\eqno{(31)}$$
with $\Phi_B(q,p,v)$ from Eqs. (11) and (12). The last terms in Eqs.
(30) and (31) (on-shell vanishes in the first order formalism) can be
neglected. Indeed, from an expression for an arbitrary variation of
$S_v$
$$
\delta S_v=-\int{\rm d}\tau\left\{\Phi_A\delta v^A+\left(\dot p_A-
{\pa L\over\pa q^A}\right)\delta q^a+\delta p_A(\dot q^A-v^A)+
\begin{array}{c}\mbox{boundary}\\ \mbox{terms}\end{array}\right\},
\eqno{(32)}$$
it follows that the transformations with parameters ${w^A}_B$ and
$f^{AB}$
$$
\begin{array}{l}\delta p_A=\Phi_B{w^B}_A,\\
\delta v^A={w^A}_B(\dot q^B-v^B);\end{array}
\eqno{(33)}$$
$$
\begin{array}{l}\delta q^A=-\Phi_Bf^{BA},\\
\delta v^A=f^{AB}\left(\dot p_A-\displaystyle{\pa L\over\pa
q^A}\right);\end{array}
\eqno{(34)}$$
are the symmetries of $S_v$ by itself. The last terms of Eqs. (30) and
(31) are precisely of this kind. Neglecting these terms, we observe
that the quantities $\Phi_B{R_a}^B$ acts as the generators of local
symmetries for $S_v$ (7).

In a similar manner, one may separating out the generators ${R_a}^A$
from the Poisson brackets in Eqs. (29), (30) and also neglect all
irrelevant terms. Then one finds symmetries $\de
q^A=\ea{R_a}^B\{q^a,\Phi_B\}$, $\de p_A=\ea
{R_a}^B\{p_a,\Phi_B\}$, accompanied by some complicated expression for
transformations of $v^A$.

The transition to the case of $S_H$ is straightforward. Indeed, note
that if $\delta S_v=0$ for arbitrary $v^i$, than, in particular,
$\delta S_v|_v=0$. Further, from Eqs. (21) and (7) we have (for
arbitrary variation of $q^A$, $p_A$, $v^\alpha$)
$$
\delta S_H=\delta(S_v|_v)=\delta S_v|_v+\Phi_i(q,p,v)|_v\delta
v^i\equiv\delta S_v|_v.
\eqno{(36)}$$

Therefore, the local symmetries for $S_H$ (21) derives from Eq. (28)
(or Eqs. (29)--(31)) by dropping $\delta v^i$ and by direct
substitution of $v^i(q^A,p_j,v^\alpha)$ in the remains. The most
transparent and symmetrical expressions for these formulae will be
obtained in the next section for the case of Lagrangian transformations
of a special form.

\section{Local symmetries of a special form}

Let us consider a special case when $\de L=0$ under the
transformations of Eq. (5). Then, in comparison with previous section,
we have an identity
$$
{\pa L\over\pa v^i}{R_a}^i(q,v)+{\pa L\over\pa v^\alpha}{R_a}^\alpha
(q,v)=0,
\eqno{(37)}$$
in addition to Eq. (27). It allows us to write the following relations
for some linear combinations of all primary constraints
$$
\Phi_\alpha(q,p){R_a}^\alpha\bigm|_v=p_A{R_a}^\alpha\bigm|_v,
\eqno{(38)}$$
$$
\Phi_\alpha(q,p){\pa{R_a}^\alpha\over\pa v^i}\biggm|_v=
p_A{\pa{R_a}^\alpha\over\pa v^i}\biggm|_v.
\eqno{(39)}$$
The first can be tested by using of Eqs. (37) and (14). To derive the
second, we differentiate the Eq. (37) with respect to $v^j$ and use the
first from Eq. (27) yield
$$
{\pa L\over\pa v^i}{\pa{R_a}^i\over\pa v^j}+{\pa L\over\pa v^\alpha}
{\pa{R_a}^\alpha\over\pa v^j}=0.
\eqno{(40)}$$
Then, substituting $v^i(q^A,p_j,v^\alpha)$ and by virtue of Eqs. (14)
and (17), we get the desired result.

Using these identities, it is not difficult to substitute
$v^i(q^A,p_j,v^\alpha)$ from Eq. (13) to the Eq. (28) or Eqs.
(29)--(31) to derive the symmetries of $S_H$ (21). We get after some
algebra
$$
\de q^A=\ea{R_a}^\alpha\bigm|_v\cdot\{q^A,\Phi_\alpha(q,p)\},
\eqno{(41)}$$
$$
\de p_A=\ea{R_a}^\alpha\bigm|_v\cdot\{p_A,\Phi_\alpha(q,p)\},
\eqno{(42)}$$
$$
\de v^\alpha=(\de q^\alpha)^.,
\eqno{(43)}$$
where ${R_a}^\alpha$ are the generators of initial Lagrangian
symmetries and $\Phi_\alpha(q,p)\approx 0$ are the all (first and
second class) primary constraints. Note that it is an exact formulae,
namely, all terms of the form (33), (34) identically cancel out
during the calculations. The equations (41)--(43) present the
Hamiltonian form of initial local Lagrangian symmetries for the system
under consideration. As in previous section, Eqs. (41) and (42) may be
identically rewritten so that the ${R_a}^\alpha$ generators are
incorporated in the Poisson bracket, neglecting trivial symmetries of
the form (33), (34).

In conclusion, note that all of the preceding remain valid for
transformations of the form $\delta q^a=\dot \epsilon^a{R_a}^A$ instead
of Eq. (5).

\section{Conclusion}

We have discussed how the local Lagrangian symmetries of an arbitrary
constrained system are transformed in passing to the first order action
$S_v$, Eq. (7), and further to the Hamiltonian action $S_H$, Eq. (21),
which includes all primary constraints.

The symmetry transformations for $S_v$ are written in the explicit form
in Eqs. (28). The formulas obtained are rewritten in terms of the
Poisson bracket to the on-shell vanishes of the first-order formalism,
Eqs. (33) and (34), and have the final form of Eqs. (29)--(31).

The explicit form of the symmetry transformations for $S_H$ are given
by Eqs. (41)--(43). Let us give some comments concerning the structure
of these equations.

(a) Equations (41) and (42) have the form similar to that of Eq. (3)
but involve a ``composite'' parameter, $\epsilon^\alpha\equiv\epsilon^q
{R_a}^\alpha\bigm|_v$. Essential parameters in the
transformations, Eqs. (41) and (42), like in the Lagrangian
transformations, Eqs. (5), are the arbitrary functions $\ea(\tau)$,
$a=1,\dots,K'$.

(b) The local symmetry generators for $S_H$ are all primary constraints
$\Phi_\alpha(q,p)$ (and not only first-order constraints, contrary to
Eq. (3)).

(c) Involved in Eqs. (41)--(43) are only the ${R_a}^\alpha$ generators
of the complete set of generators, ${R_a}^A$. It is not surprising
since in the Lagrangian formalism the generators ${R_a}^i$ can also be
expressed through the rest ones, using identity (27), as ${R_a}^i\equiv
-\tilde M^{ij}M_{ij}{R_a}^\alpha$.

It would be of interest to generalize the statements of the present
work to the case of a complete Hamiltonian action $S_c$, Eq. (2), i.e.,
to take into account all secondary constraints of the theory, and to
discuss the problem of the deformation of the algebra of original
Lagrangian transformations in the transition $S\to S_v\to S_H\to S_c$.
The work in the direction is in progress.

\section*{Acknowledgments}

This work is supported in part by ISF Grant No M21000 and European
Community Grant No INTAS-93-2058.

\end{document}